\documentclass[twocolumn,twoside,slac_two]{revtex4}
\usepackage{graphicx}
\usepackage{subfigure}
\usepackage{fancyhdr}
\begin{document}

\title{Anisotropies in the cosmic radiation observed with ARGO-YBJ}


\author{R. Iuppa}
\affiliation{Department of Physics, University of Tor Vergata and INFN, Sezione di Roma Tor Vergata, via della Ricerca Scientifica 1,  00133 Roma , Italy}
\author{on behalf of the ARGO-YBJ collaboration}
\affiliation{}

\begin{abstract}
Important informations on the origin and the propagation mechanisms of cosmic rays may be provided by the measurement of the anisotropies of their arrival direction.
In this paper the observation of anisotropy regions at different angular scales is reported. 
In particular, the observation of a possible anisotropy on scales between $\sim$10$^{\circ}$ and $\sim$30$^{\circ}$ may be a key-detection for speculations on the presence of unknown features of the magnetic fields the charged cosmic rays propagate through, as well as to potential contributions of nearby sources to the total flux of cosmic rays. 
Evidence of new weaker few-degree excesses throughout the sky region $195^{\circ}\leq$ R.A. $\leq 315^{\circ}$ is also reported.
\end{abstract}

\maketitle

\thispagestyle{fancy}


\section*{Introduction}
As the most part of cosmic rays (CRs) are charged nuclei, their arrival direction is deflected and made isotropic by the action of galactic magnetic field (GMF) that they propagate through before reaching the Earth atmosphere. In such a field, the gyroradius of CRs is given by
$r_{a.u.}=100\,R_{\textrm{\scriptsize{TV}}}$,
where $r_{a.u.}$ is in astronomic units  and $R_{\textrm{\scriptsize{TV}}}$ is in TeraVolt. It must be taken as a reference value, because the GMF is the superposition of regular field lines and chaotic contributions, the strength of both them being still under debate. Data available today gives for the local total intensity the value $B=2\div 4\textrm{ $\mu$G}$. 

Actually, deviations from the isotropy are expected to occurr as a consequence of the particular realization of the random distribution of cosmic ray sources and magnetic field lines in the galaxy \cite{blasi}. In this framework, the amplitude of the anisotropy is proportional to the rigidity, that is why it is expected to be seen at high energies (above few hundreds TeV). However, different experiments \cite{nagashima,kam07,tibet06,milagro09,eastop09,icecube11} observed an energy-dependent \emph{"large scale"} anisotropy in the sidereal time frame, well below that threshold. The amplitude is about 10$^{-4}$ - 10$^{-3}$, suggesting the existence of two distint broad regions, one showing an excess of CRs (called "tail-in"), distributed around 40$^{\circ}$ to 90$^{\circ}$ in Right Ascension (R.A.). The other a deficit (the "loss cone"), distributed around 150$^{\circ}$ to 240$^{\circ}$ in R.A..
The origin of these anisotropies is still unknown. Some authors claim that the observations may be due to a combined effect of the regular and turbolent GMF \cite{battaner09}, or to local uni- and bi-dimensional inflows \cite{amenomori10}.
Other studies suggest that it can be explained within the diffusion approximation taking into account the role of the few most nearby and recent sources \cite{blasi,erlykin06}.

Easy to understand, more beamed the anisotropies and lower their energy, more difficult to fit the standard model of CRs and GMF to experimental results. That is why the evidence of the existence of a medium angular scale anisotropy contained in the tail-in region by the Tibet AS$\gamma$ \cite{amenomori07} and Milagro \cite{milagro08} collaborations in rcent years was rather surprising. 
Similar small scale anisotropies has been recently claimed to be observed by the Icecube experiment in the Southern hemisphere \cite{icecube11}.
So far, no theory of CRs in the Galaxy exists which is able to explain few degrees anisotropies in the rigidity region 1-10 TV leaving the standard model of CRs and that of the local GMF unchanged at the same time. 

From the experimental viewpoint, observing anisotropy effects at the level of 10$^{-4}$ with an air shower array is a difficult job, because of the intrinsic difficulties that this kind of apparatus has to cope with in estimating the exposure. 

Finally, the observation of a possible small angular scale anisotropy region contained inside a larger one rely on the capability for suppressing the anisotropic structures at larger scales without, at the same time, introducing effects of the analysis on smaller scales.

In this paper the observation of CR anisotropy at different angular scales with ARGO-YBJ is reported as a function of the primary energy.
\section{The ARGO-YBJ experiment}
The ARGO-YBJ experiment, located at the YangBaJing Cosmic Ray
Laboratory (Tibet, P.R. China, 4300 m a.s.l., 606 g/cm$^2$), is an air shower array able to detect the cosmic radiation at an energy threshold of a few hundred GeV. The full detector is in stable data taking since November 2007 with a duty cycle greater than 85\%. The trigger rate at the threshold is 3.6 kHz. The detector characteristics and performance are described in \cite{moon11}.
%
%%%%%%%%%%%%%%%%%%%%%%%%%%%%%%%%%%%%%%
\section{Data analysis and Results}
%%%%%%%%%%%%%%%%%%%%%%%%%%%%%%%%%%%%%%
%
In order to study the anisotropy at different angular scales the isotropic background of CRs has been estimated with two methods: the equi-zenith angle method \cite{amenomori05} and the direct integration method \cite{Fleysher}.

The equi-zenith angle method, used to study the large scale anisotropy, is able to eliminate various spurious effects caused by instrumental and environmental variations, such as changes in pressure and temperature that are hard to control and tend to introduce systematic errors in the measurement. The method uses data coming from all the angular scales, so that potential small structures are not separated from the underlying large scale modulation.

The direct integration method, based on time-average, rely on the assumption that the local distribution of the incoming CRs is slowly varying and the time-averaged signal may be used as a good estimation of the background content. 
Time-averaging methods act effectively as a high-pass filter, not allowing to inspect features larger than the time over which the background is computed (i.e., 15$^{\circ}$/hour$\times \Delta t$ in R.A.). The time interval used to compute the average spans $\Delta t$= 3 hours and makes us confident the results are reliable for structures up to $\approx$35$^{\circ}$ wide. 

%
%%%%%%%%%%%%%%%%%%%%%%%%%%%%%%%%%%%%%%%%%%%%%%%%%%%%%%%%%%%%%%%%%%%%
%%%%%%%%%%%%%%%%%%%%%%%%%%%%%%%%%%%%%%%%%%%%%%%%%%%%%%%%%%%%%%%%%%%
%

%%\section{Results}
\subsection{Large Scale Anisotropy}
\begin{figure}[!tbp]
\centering
\includegraphics[width=0.45\textwidth]{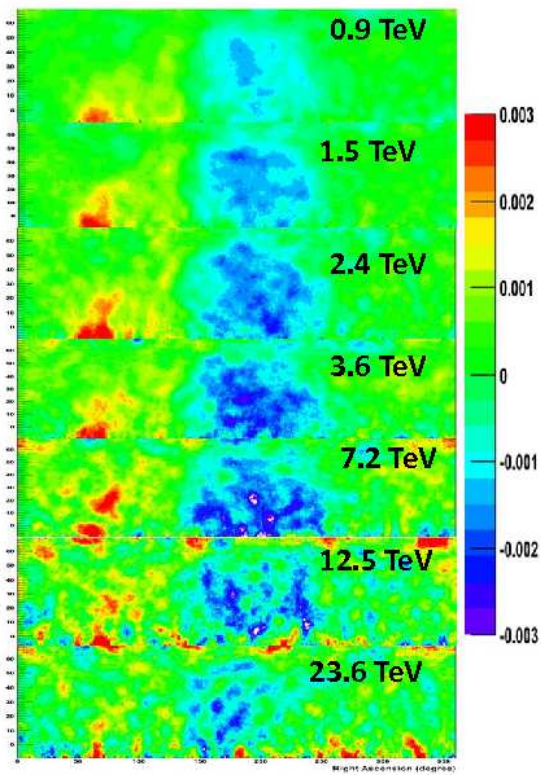}
  \caption{Large scale CR anisotropy observed by ARGO-YBJ as a function of the energy. The color scale gives the relative CR intensity.}
\label{fig:large_scale}
\end{figure}

The observation of the CR large scale anisotropy by ARGO-YBJ is shown in the figure \ref{fig:large_scale} at different primary energy up to about 25 TeV. 
The data used in this analysis was collected by ARGO-YBJ from 2008 January
to 2009 December with a reconstructed zenith angle $\leq$ 45$^{\circ}$.
The so-called \textit{`tail-in'} and \textit{`loss-cone'} regions, correlated to an enhancement and a deficit of CRs, are clearly visible with a statistical significance greater than 20 s.d..
The tail-in broad structure appears to dissolve to smaller angular scale spots with increasing energy. It should be stressed that the energies reported in the figure \ref{fig:large_scale} refer to the median energy of all nuclei triggering the experiment.

\begin{figure}[!tbp]
\centering
  \includegraphics[width=0.45\textwidth]{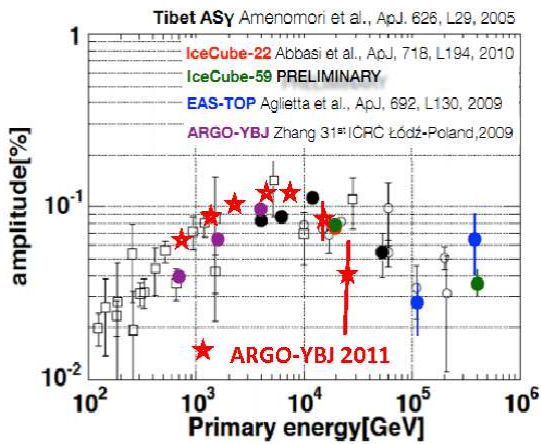}
\caption{Amplitude of the first harmonic as a function of the energy, compared to other measurements.} 
\label{fig:large_amplitude}
\end{figure}

To quantify the scale of the anisotropy we studied the 1-D R.A. projections integrating the sky maps inside a declination band given by the field of view of the detector. Therefore, we fitted the R.A. profiles with the first two harmonics. The resulting amplitude of the first harmonic is plotted in the right plot of figure \ref{fig:large_amplitude} where is compared to other measurements as a function of the energy. The ARGO-YBJ results are in agreement with other experiments suggesting a decrease of the anisotropy first harmonic amplitude with increasing energy.
%%
%%%%%%%%%%%%%%%%%%%%%%%%%%%%%%%%%%%%%%%%%%%%%%%%%%%%%%%%%%%%%%%%%%%%
\begin{figure*}[!tbp] 
\centering
\includegraphics[width=0.85\textwidth]{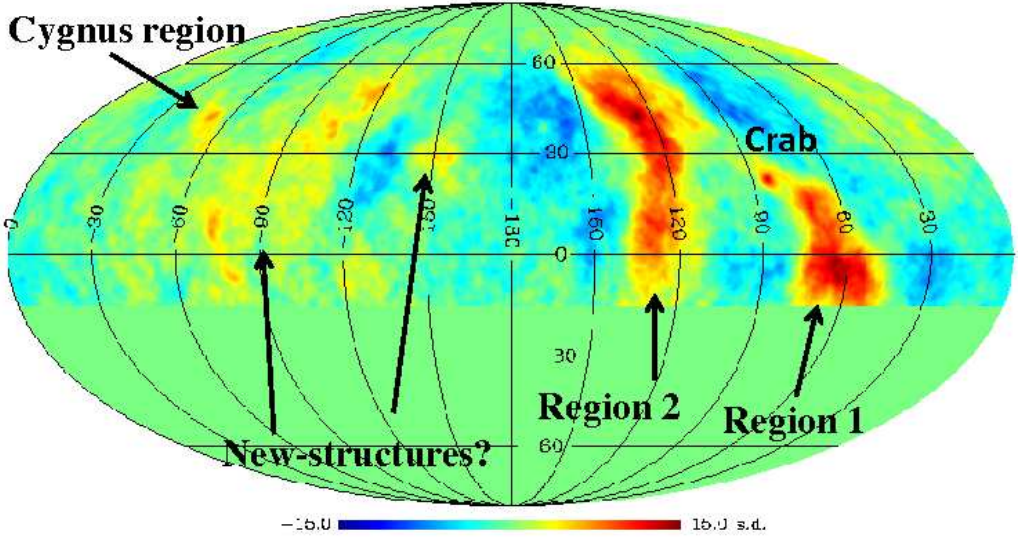}
  \caption{Intermediate scale CR anisotropy observed by ARGO-YBJ. The color scale gives the statistical significance of the observation in standard deviations.}
\label{fig:intermediate_scale}
\end{figure*}

\begin{figure}[t]
\centering
\includegraphics[width=0.45\textwidth]{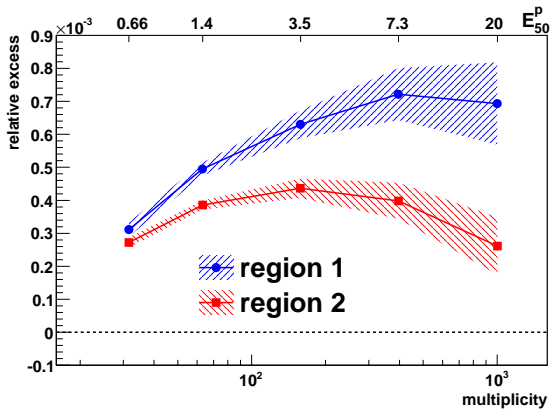}
\caption{Size spectrum of the regions 1 and 2. The vertical axis represents the ratio between the events collected. The upper scale shows the corresponding proton median energy (see text).} 
\label{fig:intermediate_amplitude}
\end{figure}
%%%%%%%%%%%%%%%%%%%%%%%%%%%%%%%%%%%%%%%%%%%%%%%%%%%%%%%%%%%%%%%%%%%%
%

\subsection{Intermediate Scale Anisotropy}
The figure \ref{fig:intermediate_scale} shows the ARGO-YBJ sky map in equatorial coordinates.
The analysis refers to events collected from November 2007 to May 2011 after the following selections: (1) $\geq$25 shower particles on the detector; (2) zenith angle of the reconstructed showers $\leq$50$^{\circ}$.
The triggering showers that passed the selection were about 2$\cdot$10$^{11}$. The zenith cut selects the declination region $\delta\sim$ -20$^{\circ}\div$ 80$^{\circ}$.
According to the simulation, the median energy of the isotropic cosmic ray proton flux is E$_p^{50}\approx$1.8 TeV (mode energy $\approx$0.7 TeV).

The most evident features are observed by ARGO-YBJ around the positions $\alpha\sim$ 120$^{\circ}$, $\delta\sim$ 40$^{\circ}$ and $\alpha\sim$ 60$^{\circ}$, $\delta\sim$ -5$^{\circ}$, positionally coincident with the regions detected by Milagro \cite{milagro08}. These regions, named ``region 1'' and ``region 2'', are observed with a statistical significance of about 14 s.d.. 
The deficit regions parallel to the excesses are due to a known effect of the analysis, that uses also the excess events to evaluate the background, artificially increasing the background.
%%If the optimal opening angle is looked for, the pre-trial maximum %%significance is 23 s.d..
On the left side of the sky map, several possible new extended features are visible, though less intense than those aforementioned.

The area $195^{\circ}\leq R.A. \leq 315^{\circ}$ seems to be full of few-degree excesses not compatible with random fluctuations (the statistical significance is more than 6 s.d. post-trial). 
The observation of these structures is reported here for the first time and together with that of regions 1 and 2 it may open the way to an interesting study of the TeV CR sky.

To figure out the energy spectrum of the excesses, data have been divided into five independent shower multiplicity sets. The number of events collected within each region are computed for the event map as well as for the background one. The ratio of these quantities is computed for each multiplicity interval. The result is shown in the figure \ref{fig:intermediate_amplitude}. Region 1 seems to have spectrum harder than isotropic CRs and a cutoff around 600 shower particles (proton median energy E$^{50}_p$ = 8 TeV). On the other hand, the excess hosted in region 2 is less intense and seems to have a spectrum more similar to that of isotropic cosmic rays.
The steepening from 100 shower particles on (E$_p^{50}$ = 2 TeV) is
likely related to efficiency effects. Further studies are on the way.
\\

\begin{figure}[t]
\centering
\includegraphics[width=0.45\textwidth]{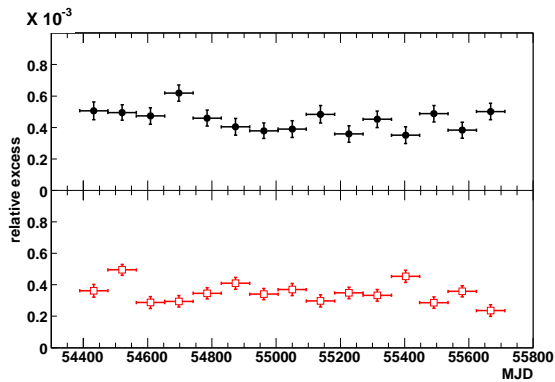}
\caption{Region 1 (upper plot) and 2 (lower plot) relative intensity as a function of UT.} 
\label{fig:time_behaviour}
\end{figure}

The figure \ref{fig:time_behaviour} reports the amplitude of the region1 and 2 anisotropies as a function of the UT. Time unit is the Modified Julian Date. It can be appreciated that no time evolution of the anisotropies are there, nor evidence of correlation of the emission.

\section{Conclusions}
This paper reports the observation of CR anisotropy at different angular scales with ARGO-YBJ, as a function of the primary energy.
The large scale CR anisotropy has been clearly observed up to about 25 TeV.
Evidence of existence of different few-degree excesses in the Northern sky (the strongest ones positionally coincident with the regions detected by Milagro in 2008) is reported. The time distribution of the phenomenon has been showed too.
A discussion of the systematic effects which may come from imprecisions in estimating the reference level is outlined.

In fact, a wrong estimation affect the significance and the relative intensity sky maps, even creating artifacts (i.e. fake excesses or deficit regions).
Drifts in detector operating conditions or atmospheric effects on the air shower development are quite hard to be modeled, then difficult to be accounted for to sufficient accuracy. As anisotropies of the order 10$^{-4}$ are looked for,
operating conditions must be known down to this level, all across the field of view and during all the acquisition time.

Background methods applied in this analysis demonstrated to be effective for analysis of diffuse from even weaker regions (e.g. see \cite{mrk421}). 

Given the importance of the topic, a joint analysis of concurrent data recorded by different experiments in both hemispheres, as well as a correlation with other observables like the interstellar energetic neutral atoms distribution \cite{ibex09,ibex11}, should be a high priority to clarify the observations.
\bigskip % extra skip inserted

\end{document}